\documentclass[%
reprint, 
twocolumn, 
amsmath,
amssymb,
aps, 
floatfix, 
superscriptaddress,
prb
]{revtex4-2}
\usepackage{dcolumn}
\usepackage{bm,graphicx,url,epsf,color}
\usepackage{amsthm}
\usepackage[colorlinks=true,linkcolor=red,citecolor=blue]{hyperref}
\usepackage{comment}
\usepackage{physics}
\usepackage{cleveref}

\newcommand{\bk}{{\bm k}}
\newcommand{\bR}{{\bm R}}
\newcommand{\br}{{\bm r}}

\newcommand{\cinam}{CNRS/Aix-Marseille Universit\'e, Centre Interdisciplinaire de Nanoscience de Marseille UMR 7325 Campus de Luminy, 13288 Marseille Cedex 9, France}

\begin{document}
\preprint{APS/123-QED}

\author{Yuncheng Mao}
\email{catmyc@gmail.com}
\affiliation{\cinam}
\author{Claudio Attaccalite}
\affiliation{\cinam}

\title{Magnetic Bulk Photovoltaic Effect in Bernal Bilayer Graphene}

\begin{abstract}
We investigate the shift-current response of inversion-broken AB-stacked bilayer graphene under in-plane and perpendicular magnetic fields, from the perturbative regime to strong orbital quantization, in both two-dimensional bulk systems and finite nanoribbons. 
An in-plane field enters through opposite momentum shifts in the two layers and leaves the bulk band dispersion essentially unchanged, producing only a modest, frequency-dependent redistribution of the shift-current spectrum. 
A weak perpendicular field is treated using a gauge-covariant Peierls expansion. 
The resulting band corrections are concentrated near Berry-curvature hot spots, although the dominant shift-current transitions occur elsewhere, and the valley-summed response is even in the field with a leading quadratic correction. At strong perpendicular field, a rational-flux magnetic
supercell reveals nearly flat Landau-level-like bulk bands and a greatly enhanced density of states, yet the bulk shift current is almost completely quenched because the relevant optical matrix elements and shift-vector contributions are suppressed or cancel. 
Finite ribbons retain optically active boundary channels: in zigzag ribbons, magnetic reconstruction turns edge-derived states from dark states into bright photovoltaic channels, with the associated peak scaling inversely with ribbon width. 
These results show that magnetic control of the nonlinear photovoltaic response is governed by wave-function reconstruction and quantum-geometric matrix elements rather than by the density of states alone.
\end{abstract}

\date{July 26, 2026}

\maketitle

\section{Introduction}

Conventional photovoltaic devices operate based on the pn-junction, where an internal electric field separates electron-hole pairs created by light absorption, driving a current through the circuit. 
The bulk photovoltaic effect (BPVE) offers a fundamentally different approach. As a second-order nonlinear optical response, BPVE generates a steady photocurrent directly in the \emph{bulk} of a non-centrosymmetric material under uniform illumination. Rather than relying on a junction's field, the current in BPVE is driven by the intrinsic crystal symmetry that enables asymmetric electron excitation and relaxation. 
This unique property allows BPVE to produce photovoltages exceeding the band gap~\cite{yang2010above,yang2017enhancement}. 
Many believed this could offer a promising route to circumvent the Shockley-Queisser limit and achieve higher light-conversion efficiencies~\cite{tan2016shift,cook2017design}. However, the analyses of Pusch \textit{et al}~\cite{Pusch2023} show that stringent limitations due to the photon energy conversion mechanism still exist.

The absence of the centrosymmetry lies at the center of the mechanism of BPVE current~\cite{Kraut1979,belinicher1980photogalvanic, balz1981theory, fridkin2001bulk}. 
This is one of the signs that BPVE is sensitive to the symmetrical properties of the physical system.
Some examples of the non-centrosymmetric materials as good candidates are the transition metal dichalcogenides (TMD)~\cite{manzeli20172d, mao2025shiftcurrent2djanus, Osterhoudt2019, He2021b, Huang2022} and the moir\'e materials~\cite{mao2025moire,atlam2025giant, Pantaleon2021, Duan2022, Huang2022, Sinha2022, Zhang2022, Huang2023, joya2025shift, Kaplan2022, Hu2023, caluga2021tstg}. 
The non-centrosymmetry can also be realized by changing the stacking manner of 2D materials without twist ~\cite{Chen2024, zheng2023gate, Postlewaite2024} or with hetero-junctions of originally centrosymmetric materials~\cite{gao2025bulk}.
Other materials with various compositions are also reported~\cite{Young2012, Tan2016, Zhang2018, Zhang2019, Xu2020, He2021, Tiwari2022, feng2025high, yang2018divergent}. 

Two major mechanisms contribute to the second-order DC current: shift current
and ballistic (or injection) current~\cite{Belinicher01011988,sipe2000second-order,parker2019diagrammatic}.
For the linearly polarized light considered here, we focus exclusively on the
shift current~\cite{balz1981theory,sipe2000second-order,tan2016shift}.
The application of the magnetic field to a physical system introduces many subtleties to the system. The external field may reduce the crystalline symmetry, and more importantly, breaks the TRS. 
BPVE under the magnetic field (therefore in the absence of TRS) is so far an under-investigated topic. Recent works show that the shift current is steady under the magnetic field, as long as the external field does not alter substantially the band structures~\cite{zhang2019switchable,dai2023magnetic}.
This scenario is referred to as the ``weak-field regime'' throughout this paper, as long as the modification to the band structure is minor and barely visible.
The results of Dai \textit{et al.}~\cite{dai2023magnetic} show that the variation of the shift current has an interesting parabolic relation with the external field in the square-lattice model at certain photon frequencies. 
We investigate how this shift-current response evolves under in-plane and
perpendicular magnetic fields in inversion-broken AB-stacked bilayer
graphene.

The paper is organized as follows:
In Section~\ref{sec: model}, the AB-bilayer graphene model is explained with all the parameters specified. We then show the formulae and necessary ingredients for the computation of the shift-current conductivity tensor.  
In Section~\ref{sec: results}, we show the results and in-depth analyses of the shift current and how it is influenced by the external field. This is followed by our statement of the main conclusions in Section~\ref{sec: conclusion}. 

\section{Theory and models}

\subsection{General considerations}

For analytical purpose, let us first consider the minimal model of AB-stacked bilayer graphene placed in magnetic fields: 
\begin{equation}\label{eq:simple ham}
\begin{split}
    & H_\xi(\br; \bm B) = \\
    & \begin{bmatrix}
        \xi v_F (\hbar \hat \bk - e\bm A_t) \cdot \bm \sigma_\xi + \Delta  & T \\
        T^\dag & \xi v_F (\hbar \hat \bk - e\bm A_b) \cdot \bm \sigma_\xi - \Delta
    \end{bmatrix}
\end{split}
\end{equation}
where 
\[
T = \begin{bmatrix}
    0 & 0 \\
    t & 0
\end{bmatrix} ,
\]
and $\xi = \pm 1$ denoting the $K$  and $K'$ valleys of graphene monolayer, respectively. $\bm \sigma_{+} = \bm \sigma =  [\sigma_x, \sigma_y]$ and $\bm \sigma_{-} = \bm \sigma^* = [\sigma_x, -\sigma_y]$, where $\sigma_{x/y/z}$ are the Pauli matrices defined on the sublattice degree of freedom of graphene.
$\Delta$ is the layer offset potential, and $\hat \bk = -i \bm \nabla$. $v_F$ is the Fermi velocity of graphene single layer. These are the minimal ingredients that capture the characteristic Mexican-hat shaped middle bands and the parabola-like higher bands of AB-bilayer graphene near the Fermi level.
$\bm A_t$ and $\bm A_b$ designate the vector potential on the top and bottom layers, respectively. In the case of penpendicular magnetic field, $\bm A_t = \bm A_b = \bm A(\br)$, whereas for in-plane parrallel magnetic field, we choose $\bm A = z \bm B \times \bm u_z$ with $\bm u_z$ being the unit vector perpendicular to the graphene planes. This results in $\bm A_t = -\bm A_b = \bm A$. The vector potentials in both layers remain uniform in-plane vector fields.  

The intersublattice-interlayer hopping breaks the centrosymmetry and is essential for the generation of second order nonlinear response. In the absence of the magnetic field, the time-reversal symmetry (TRS) is present by satisfying $H^*_\xi = H_{-\xi}$. Non-zero magnetic field breaks the TRS, but the following ration holds:
\begin{equation}
    H^*_\xi(\bm B) = H_{-\xi}(-\bm B) .
\end{equation}
This relation suggests that reversing the orientation of the magnetic field leaves the energy spectrum of the system unchanged, up to an interchange of valley states. This fixes the relations of the matrix elements of velocity operator and \emph{interband} connections for $n \neq m$: 
\begin{eqnarray}
    & \bm v_{nm,\xi} (\bk, \bm B) = -\bm v_{mn,-\xi}(-\bk, -\bm B) ; \\
    & \bm r_{nm; \xi} (\bk, \bm B) = \bm r_{mn,-\xi}(-\bk, -\bm B) .
\end{eqnarray}
The \emph{intraband} Berry connection
\[\bm{\mathcal{A}}_{nn,\xi}(\bk,\bm B) = i \mel{u_{nk,\xi}(\bm B)}{\partial_{k_a}}{u_{nk,\xi}(\bm B)} \]
also validates
\begin{equation}
 \bm{\mathcal{A}}_{nn,\xi}(\bk,\bm B) = \bm{\mathcal{A}}_{nn,\xi}(-\bk,-\bm B)
\end{equation}
up to a gauge change. 
This leads to the shift vector being TRS \emph{even}:
\begin{equation}
    R^{a,b}_{nm,\xi}(\bk, \bm B) = R^{a,b}_{nm,-\xi}(-\bk, -\bm B)
\end{equation}
with
\begin{equation}
    \begin{split}
    & R^{a,b}_{nm,\xi}(\bk, \bm B) =\\
    & \mathcal{A}^a_{nn,\xi}(\bk, \bm B) - \mathcal{A}^a_{mm,\xi}(\bk,\bm B) - \partial_{a} \mathrm{Arg}(r^{b}_{nm,\xi}(\bk)) ,
    \end{split}
\end{equation}
where $a,b \in \{x, y, z\}$. 
The shift-current conductivity is
\begin{equation}\label{eq:sc}
\begin{split}
    & \sigma^{abc}(\omega)  =  - \frac{i \pi e^3}{\hbar^2} \int \frac{\dd^D \bk}{(2\pi)^D} \times \\ 
    & \sum_{nm,\xi} f_{nm} \left[ r^b_{mn} r^c_{nm;a} + r^c_{mn} r^b_{nm;a} \right] \delta(\omega_{mn} - \omega),
\end{split}
\end{equation}
Here the dependence on $\bk$, $\bm B$, and valley is hidden for simplicity.
It is then direct to identify that shift current is \emph{even} function of the magnetic field:
\[
\sigma^{abc}(\omega, \bm B) = \sigma^{abc}(\omega, -\bm B),
\]
given that 
\[
r^{b}_{nm;a} = -i R^{a,b}_{nm} r^b_{nm}.
\]
Consequently, a smooth expansion of the valley-summed shift-current
conductivity around $\bm B=\bm0$ has no linear term and begins at order
$B^2$.
The reasoning here can be easily generalized to more detailed model description in Sec.~\ref{sec: model}.

\subsection{Unperturbed connections and shift vectors}
We first explore the expressions of the connections and shift vectors in the absence of magnetic field. 
For simplicity, we focus on the $K$-valley first. The physical quantities from the $K'$-valley can be inferred from time-reversal operation on the wave functions.

As the two middle bands contribute the strongest signal to the nonlinear optical response, we only look at the interband connections, shift vectors defined on the middle bands. 
Their energies are, in the absence of magnetic field,  given by
\begin{equation}
    E_s = s ~ \varepsilon(\bk), 
\end{equation}
where the subscript $s = \pm$ denotes the conduction and valence bands, respectively, and 
\[
 \varepsilon(\bk) = \sqrt{\Delta^2 + \rho^2 + \frac{t^2}{2} - \sqrt{\frac{t^4}{4} + \rho^2 (t^2 + 4 \Delta^2) }} ,
\]
with $\rho = \hbar v_F |\bk|$.
The spinor wave funcitons write 
\begin{equation}\label{eq: wf analytical}
    \ket{u_{s\bk}} = \mathcal{N}_{s} \begin{pmatrix}
        t \rho (E_s+\Delta)e^{-i\varphi} \\
        t(E_s^2 - \Delta^2) \\
        [(E_s^2 - \Delta^2)-\rho^2](E_s+\Delta)\\
        \rho [(E_s-\Delta)^2 -\rho^2]e^{i\varphi}
    \end{pmatrix}, 
\end{equation}
where $\mathcal{N}_s$ is the normalizing factor for band $s$. Note that form of the wave function already implies a choice of gauge. The angular phase $\varphi$ is related to $\bk$ such that $k_x + i k_y = |\bk| e^{i\varphi}$.
For simplicity, we note the wave functions as 
\[
\ket{u_{s\bk}} = \begin{pmatrix}
    C^s_1(k) e^{-i\varphi} \\
    C^s_2(k) \\
    C^s_3(k) \\
    C^s_4(k) e^{i\varphi}
\end{pmatrix}, s = \pm.
\]
Here $k = |\bk|$, and $C^s_\mu(k)$ with $\mu = 1, 2, 3, 4$ are all \emph{real} coefficients. 
It is then convenient to define the two following quantities: 
\begin{equation}
    U(k) = \frac{C^+_1C^-_1-C^+_4C^-_4}{k},
\end{equation}
and 
\begin{equation}
V(k)=\sum_{\mu=1}^4 C^+_\mu(k)\frac{d C^-_\mu(k)}{dk}.
\end{equation}
With the gauge chosen, the interband connection can be conveniently expressed as:
\begin{equation}
    r^x_{vc} = - U \sin \varphi - i V \cos \varphi, 
\end{equation}
and 
\begin{equation}
    r^y_{vc} = U \cos \varphi - i V \sin \varphi .
\end{equation}
The \emph{intraband} Berry connection difference is essential for the computation of shift vector:
\begin{equation}
    \bm A_v - \bm A_c = \frac{\Lambda(k)}{k} \mathbf{e}_\varphi,
\end{equation}
where $\mathbf{e}_\varphi$ is the tangent unit vector, and 
\begin{equation}
    \Lambda(k) = \left[(C^-_1)^2-(C^-_4)^2\right] 
-\left[(C^+_1)^2-(C^+_4)^2\right].
\end{equation}
The analytical form of the shift vector is given by 
\begin{equation}    
R^{a;b}_{vc}
=
\frac{\Lambda(k)}{k} \hat{\bm \varphi}_a
-
\partial_{k_a}\arg r^b_{vc}
,
\quad a,b\in\{x,y\}.
\end{equation}
More specifically, we have
\begin{eqnarray}
R^{x;x}_{vc}
=
-\frac{\Lambda\sin\varphi}{k}
-
\frac{\sin\varphi}{D_x}
\left[
W\cos^2\varphi+\frac{UV}{k}
\right] ,
\\
R^{x;y}_{vc}
=
-\frac{\Lambda\sin\varphi}{k}
+
\frac{\sin\varphi}{D_y}
\left[
W\cos^2\varphi-\frac{UV}{k}
\right],
\\
R^{y;x}_{vc}
=
\frac{\Lambda\cos\varphi}{k}
-
\frac{\cos \varphi}{D_x} \left[W\sin^2\varphi
-
\frac{UV}{k}\right], 
\end{eqnarray}
and 
\begin{equation}
R^{y;y}_{vc}
=
\frac{\Lambda\cos\varphi}{k}
+
\frac{\cos\varphi}{D_y}
\left[
W\sin^2\varphi+\frac{UV}{k}
\right],
\end{equation}
where 
\begin{equation}
D_x=U^2\sin^2\varphi+V^2\cos^2\varphi,
\end{equation}
\begin{equation}
D_y=U^2\cos^2\varphi+V^2\sin^2\varphi,
\end{equation}
\begin{equation}
W=U\partial_{k}V-V \partial_{k}U .
\end{equation}

\subsection{Shift current perturbed by in-plane magnetic field}

Denoting $\bm \kappa(\bm B) = e\bm A/\hbar$, it is clear that the in-plane magnetic field shifts the wave vectors of the Bloch states of each layer in opposite directions. The magnitude of $\bm \kappa$ is directly proportional to the intensity of the magnetic field but its orientation is orthogonal to the latter. 
Eq.\eqref{eq:simple ham} becomes 
\begin{equation}\label{eq:simple inplane}
\begin{split}
    & H_\xi(\bk; \bm B) =\\
    & \begin{bmatrix}
        \xi \hbar v_F (\bk - \bm\kappa) \cdot \bm \sigma_\xi + \Delta  & T \\
        T^\dag &  \xi \hbar v_F ( \bk + \bm\kappa) \cdot \bm \sigma_\xi - \Delta
    \end{bmatrix} .
\end{split}
\end{equation}
In the following analytical discussion we focus on the $K$ valley. The $K'$ valley is simply related to the $K$ valley via time-reversal operation.
For weak in-plane magnetic field, Eq.~\eqref{eq:simple inplane} can be written as
\begin{equation}
    H(\bk,\bm\kappa)=H_0(\bk)+\delta H_{\parallel},
    \qquad
    \delta H_{\parallel}=\hbar \sum_{j=x,y}\kappa_j \hat{\Xi}_j ,
\end{equation}
with the layer-odd velocity operator
\begin{equation}
    \hat{\Xi}_j =
    v_F
    \begin{bmatrix}
        -\sigma_j & 0\\
        0 & \sigma_j
    \end{bmatrix}.
\end{equation}
The ordinary velocity operator entering the optical transition is instead
\begin{equation}
    \hat v^b
    =
    \frac{1}{\hbar}\partial_{k_b}H_0(\bk)
    =
    v_F
    \begin{bmatrix}
        \sigma_b & 0\\
        0 & \sigma_b
    \end{bmatrix},
    \qquad b=x,y .
\end{equation}
Thus the magnetic perturbation is generated by a layer-odd velocity, whereas the optical matrix element is generated by the layer-even velocity.

Let $n$ denote one of the unperturbed bands and $\ell$ all other bands.
To first order in $\bm\kappa$, the perturbed wave function is
\begin{equation}
    \ket{u_n(\bm\kappa)}
    =
    \ket{u_n}
    +
    \kappa_j 
    \sum_{\ell\neq n}
    \ket{u_\ell}
    \frac{\Xi^j_{\ell n}}{\omega_{n\ell}}
    +\mathcal{O}(\kappa^2),
\end{equation}
where $\omega_{n \ell} = (E_n - E_\ell)/\hbar$ and repeated Cartesian index $j=x,y$ is summed silently. We also have
\[
    \Xi^j_{\ell n}=\mel{u_\ell}{\hat{\Xi}_j}{u_n}.
\]
The transition frequency between the two middle bands changes as
\begin{equation}
    \omega_{cv}(\bm\kappa)
    =
    \omega_{cv}
    +
    \kappa_j \Delta \Xi^j_{cv}
    +\mathcal{O}(\kappa^2),
    \qquad
    \Delta \Xi^j_{cv}=\Xi^j_{cc}-\Xi^j_{vv},
\end{equation}
where $\omega_{cv}=(E_c-E_v)/\hbar>0$.

Since the interband connection $r^b_{vc}(\bm \kappa)$ is related to the velocity matrix elements by $r^b_{vc} = v^b_{vc}/i\omega_{vc}$, we first look into 
\begin{equation}
    v^b_{vc}(\bm\kappa)
    =
    \mel{u_v(\bm\kappa)}{\hat v^b}{u_c(\bm\kappa)} .
\end{equation}
Since $\delta H_{\parallel}$ is independent of $\bk$, the velocity operator itself has no first-order correction in this minimal continuum model. 
The correction to $v^b_{vc}$ therefore comes only from the wave functions:
\begin{equation}
    v^b_{vc}(\bm\kappa)
    =
    v^b_{vc}
    +
    \hbar\kappa_j \mathcal{L}^{b,j}_{vc}
    +
    \mathcal{O}(\kappa^2),
\end{equation}
with
\begin{equation}
    \mathcal{L}^{b,j}_{vc}
    =
    \sum_{\ell\neq v}
    \frac{\Xi^j_{v\ell}v^b_{\ell c}}{E_v-E_\ell}
    +
    \sum_{\ell\neq c}
    \frac{v^b_{v\ell}\Xi^j_{\ell c}}{E_c-E_\ell}.
\end{equation}
The interband connection for the $v\rightarrow c$ transition is then
\begin{equation}
    r^b_{vc}(\bm\kappa)
    =
    i\frac{v^b_{vc}(\bm\kappa)}{\omega_{cv}(\bm\kappa)}
    =
    r^b_{vc}
    +
    \kappa_j r^{b,j}_{vc}
    +
    \mathcal{O}(\kappa^2),
\end{equation}
where
\begin{equation}
    r^{b,j}_{vc}
    =
    i\frac{\hbar \mathcal{L}^{b,j}_{vc}}{\omega_{cv}}
    -
    i\frac{v^b_{vc}}{\omega_{cv}^2}\Delta \Xi^j_{cv}.
\end{equation}
The second term changes the magnitude of the interband connection through the shift of the transition frequency. 
Because $\omega_{cv}(\bm\kappa)$ is real, this term does not by itself introduce a new optical phase.

The shift vector follows from
\begin{equation}
    R^{a;b}_{vc}(\bm\kappa)
    =
    A^a_v(\bm\kappa)-A^a_c(\bm\kappa)
    -
    \partial_{k_a}\arg v^b_{vc}(\bm\kappa),
\end{equation}
where the phase of $v^b_{vc}$ can be used instead of the phase of $r^b_{vc}$ because the two differ only by the real transition frequency and a constant phase factor.
The intraband Berry connections have the expansion
\begin{equation}
    A^a_n(\bm\kappa)
    =
    A^a_n
    +
    \kappa_j \mathcal{A}^{a,j}_n
    +
    \mathcal{O}(\kappa^2),
\end{equation}
with
\begin{equation}
    \mathcal{A}^{a,j}_n
    =
    \hbar
    \sum_{\ell\neq n}
    \frac{
    \Xi^j_{n\ell}r^a_{\ell n}
    +
    \Xi^j_{\ell n}r^a_{n\ell}
    }{E_n-E_\ell}.
\end{equation}
Consequently,
\begin{equation}
    R^{a;b}_{vc}(\bm\kappa)
    =
    R^{a;b}_{vc}
    +
    \kappa_j \mathcal{R}^{a;b,j}_{vc}
    +
    \mathcal{O}(\kappa^2),
\end{equation}
where
\begin{equation}
    \mathcal{R}^{a;b,j}_{vc}
    =
    \mathcal{A}^{a,j}_v
    -
    \mathcal{A}^{a,j}_c
    -
    \partial_{k_a}
    \Im\left[
        \frac{\hbar \mathcal{L}^{b,j}_{vc}}{v^b_{vc}}
    \right].
\end{equation}
This expression makes explicit that the in-plane magnetic field changes the shift vector through the field-induced mixing of the zero-field bands.

For linearly polarized light along $b$, the contribution of the two middle bands to the shift current can be written as
\begin{equation}
\begin{split}
    \sigma^{abb}_{vc}(\omega,\bm\kappa)
    ={}&
    -\frac{2\pi e^3}{\hbar^2}
    \int\frac{\dd^2\bk}{(2\pi)^2}
    f_{vc}
    \\
    &\times
    \frac{|v^b_{vc}(\bm\kappa)|^2}{\omega_{cv}^2(\bm\kappa)}
    R^{a;b}_{vc}(\bm\kappa)
    \delta\left[\omega-\omega_{cv}(\bm\kappa)\right].
\end{split}
\end{equation}
Expanding the integrand to linear order,
\[
    \mathcal{I}^{a;b}_{vc}(\omega,\bk,\bm\kappa)
    =
    \mathcal{I}^{a;b}_{vc}
    +
    \kappa_j \mathcal{I}^{a;b,j}_{vc}
    +\mathcal{O}(\kappa^2),
\]
shows three distinct perturbative effects. Here
\[
    \mathcal{I}^{a;b}_{vc}
    =
    \frac{|v^b_{vc}|^2}{\omega_{cv}^2}
    R^{a;b}_{vc}
    \delta_0 ,
    \qquad
    \delta_0=\delta(\omega-\omega_{cv}),
\]
and the first-order coefficient is
\begin{equation}
\begin{split}
    \mathcal{I}^{a;b,j}_{vc}
    =
    &\left[
    \frac{2\hbar}{\omega_{cv}^2}
    \Re\left[(v^b_{vc})^*\mathcal{L}^{b,j}_{vc}\right]
    -
    \frac{2|v^b_{vc}|^2\Delta\Xi^j_{cv}}{\omega_{cv}^3}
    \right]
    R^{a;b}_{vc}\delta_0\\
    &+
    \frac{|v^b_{vc}|^2}{\omega_{cv}^2}
    \mathcal{R}^{a;b,j}_{vc}\delta_0
    -
    \frac{|v^b_{vc}|^2}{\omega_{cv}^2}
    R^{a;b}_{vc}
    \Delta\Xi^j_{cv}\delta'_0 ,
\end{split}
\end{equation}
where $\delta'_0 = \dd \delta(\omega - \omega_{cv})/\dd \omega$ .
The first line describes the change of optical transition strength, the second term describes the change of the shift vector itself, and the final term describes the displacement of spectral weight caused by the shift of the transition energy.
For one valley this correction is generally linear in $\bm\kappa$ and anisotropic in the angle between $\bk$ and $\bm\kappa$.
After summing equally occupied $K$ and $K'$ valleys, however, time reversal requires the total shift current to be an even function of the magnetic field.
Thus the valley-summed shift current has no linear correction in weak
in-plane magnetic field, although a single-valley contribution can retain a
linear-in-field term.

\subsection{Shift current in weak perpendicular magnetic field}
\label{subsec:perpendicular-weak-field}

A weak perpendicular field cannot be treated as a simple momentum shift, because the Peierls phase through each plaquette is finite and ordinary two-dimensional crystal momentum is no longer a good quantum number. 
Nevertheless, before the Landau levels are spectrally resolved, the response can be expanded around the zero-field Bloch problem by using gauge-covariant derivatives in momentum space~\cite{Luttinger1955motion, panati2003effective, gosselin2008semiclassical}. 
This gives an analytic weak-field counterpart of the Landau-level description.

Let $H_0(\bk)\ket{u_n}=E_n\ket{u_n}$ be the zero-field Bloch problem. For a
matrix $O_{nm}(\bk)$ written in the band basis, we define the covariant
derivative
\begin{equation}
    D_a O = \partial_{k_a}O - i[A^a,O],
    \qquad
    A^a_{nm}=i\mel{u_n}{\partial_{k_a}}{u_m},
\end{equation}
where $a=x,y$ and $[X, Y] = XY - YX$ is the commutator of operators $X$ and $Y$. Under a $\bk$-dependent phase rotation of the bands, $D_aO$
transforms in the same way as $O$ itself. The position operator projected to
the Bloch basis is therefore represented by $iD_a$.

In symmetric gauge, the magnetic field enters through the gauge-covariant
Peierls expansion
\begin{equation}\label{eq: ham weak perp B}
    H(B_z)=H_0+B_z\mathcal{M}^z+\mathcal{O}(B_z^2),
\end{equation}
with
\begin{equation}
    \mathcal{M}^z
    =
    \frac{e}{4\hbar}
    \left[
    \left\{\partial_{k_y}H_0,iD_x\right\}
    -
    \left\{\partial_{k_x}H_0,iD_y\right\}
    \right],
\end{equation}
where the sign convention follows the Peierls phase used later in the tight-binding Hamiltonian in Sec.~\ref{sec: model}. 
$\{X, Y\} = XY + YX$ is the anticommutator of operators $X$ and $Y$.
The derivation from the real-space Peierls phase to this
band-basis expression is given in Sec.~S2
of the Supplemental Material~\cite{supplemental_material}.
\nocite{bernevig2013topological,cheng2017nonlinearLL,mccann2006landau,mccann2013electronic}
The diagonal matrix element is the orbital magnetic correction to the band energy; 
equivalently, $\mathcal{M}^z_{nn}=-m^z_n$ if $m^z_n$ denotes
the usual orbital magnetic moment. Choosing the first-order correction to be orthogonal to $\ket{u_n}$, the band energy and wave function become
\begin{equation}
    E_n(B_z)=E_n+B_z\mathcal{M}^z_{nn}
    +\mathcal{O}(B_z^2),
\end{equation}
\begin{equation}
    \ket{u_n(B_z)}
    =
    \ket{u_n}
    +
    B_z\sum_{\ell\neq n}
    \eta^z_{\ell n}\ket{u_\ell}
    +\mathcal{O}(B_z^2),
\end{equation}
with 
\[
\eta^z_{\ell n} =
    \frac{\mathcal{M}^z_{\ell n}}{E_n-E_\ell}.
\]
Consequently, 
\begin{equation}
    \omega_{mn}(B_z)
    =
    \omega_{mn}
    +
    B_z\Omega^z_{mn}
    +
    \mathcal{O}(B_z^2) ,
\end{equation}
where 
\[
 \Omega^z_{mn} =
    \frac{\mathcal{M}^z_{mm}-\mathcal{M}^z_{nn}}{\hbar}.
\]

For a transition $n\rightarrow m$, with
$\omega_{mn}=(E_m-E_n)/\hbar>0$, the transition frequency is corrected by
$\Omega^z_{mn}$. The optical velocity matrix element has the expansion
\begin{equation}
    v^b_{nm}(B_z)
    =
    v^b_{nm}
    +
    B_z\mathcal{V}^{b,z}_{nm}
    +
    \mathcal{O}(B_z^2),
\end{equation}
where
\begin{equation}
\begin{split}
    \mathcal{V}^{b,z}_{nm}
    ={}&
    \frac{1}{\hbar}(D_b\mathcal{M}^z)_{nm}
    +
    \sum_{\ell\neq n}
    (\eta^z_{\ell n})^* v^b_{\ell m}
    \\
    &+
    \sum_{\ell\neq m}
    v^b_{n\ell}\eta^z_{\ell m}.
\end{split}
\end{equation}
The first term is the explicit correction to the velocity operator, while the
last two terms are generated by the magnetic correction to the wave functions.
For non-degenerate optical transitions,
\begin{equation}
    r^b_{nm}(B_z)
    =
    \frac{v^b_{nm}(B_z)}{i \omega_{nm}(B_z)}
    =
    r^b_{nm}
    +
    B_z r^{b,z}_{nm}
    +
    \mathcal{O}(B_z^2),
\end{equation}
with
\begin{equation}
    r^{b,z}_{nm}
    =
    \frac{\mathcal{V}^{b,z}_{nm}}{i \omega_{nm}}
    +
    \frac{v^b_{nm}}{i \omega_{nm}^2}\Omega^z_{mn}.
\end{equation}

The intraband Berry connection is corrected as
\begin{equation}
    A^a_n(B_z)=A^a_n+B_z\mathcal{A}^{a,z}_n+\mathcal{O}(B_z^2),
\end{equation}
where
\begin{equation}
    \mathcal{A}^{a,z}_n
    =
    \sum_{\ell\neq n}
    \left[
    (\eta^z_{\ell n})^*r^a_{\ell n}
    +
    \eta^z_{\ell n}r^a_{n\ell}
    \right].
\end{equation}
Consequently, the shift vector for the same transition satisfies
\begin{equation}
    R^{a;b}_{nm}(B_z)
    =
    R^{a;b}_{nm}
    +
    B_z\mathcal{R}^{a;b,z}_{nm}
    +
    \mathcal{O}(B_z^2),
\end{equation}
with
\begin{equation}
    \mathcal{R}^{a;b,z}_{nm}
    =
    \mathcal{A}^{a,z}_n
    -
    \mathcal{A}^{a,z}_m
    -
    \partial_{k_a}
    \Im\left[
        \frac{r^{b,z}_{nm}}{r^b_{nm}}
    \right].
\end{equation}
This is the perpendicular-field analogue of the in-plane correction above, but with the magnetic perturbation generated by the \emph{covariant} real-space position operators rather than by a layer-dependent momentum shift.

For linearly polarized light along $b$, define
\begin{equation}
    Q^{a;b}_{nm}
    =
    \frac{|v^b_{nm}|^2}{\omega_{mn}^2}
    R^{a;b}_{nm}.
\end{equation}
Its first-order magnetic correction is
\begin{equation}
\begin{split}
    Q^{a;b,z}_{nm}
    =
    &
    \left[
    \frac{2\Re\left[(v^b_{nm})^*\mathcal{V}^{b,z}_{nm}\right]}
    {\omega_{mn}^2}
    -
    \frac{2|v^b_{nm}|^2\Omega^z_{mn}}
    {\omega_{mn}^3}
    \right]
    R^{a;b}_{nm}
    \\
    &+
    \frac{|v^b_{nm}|^2}{\omega_{mn}^2}
    \mathcal{R}^{a;b,z}_{nm}.
\end{split}
\end{equation}
The shift-current tensor then takes the weak-field form
\begin{equation}
    \sigma^{a;bb}(\omega,B_z)
    =
    \sigma^{a;bb}_{(0)}(\omega)
    +
    B_z\sigma^{a;bb}_{(1)}(\omega)
    +
    \mathcal{O}(B_z^2),
\end{equation}
where
\begin{equation}
\begin{split}
    \sigma^{a;bb}_{(1)}(\omega)
    =
    -\frac{2\pi e^3}{\hbar^2}
    \sum_{nm}
    \int\frac{\dd^2\bk}{(2\pi)^2}
    \Big[
    &f^z_{nm}Q^{a;b}_{nm}\delta_0
    +
    f_{nm}Q^{a;b,z}_{nm}\delta_0
    \\
    &-
    f_{nm}Q^{a;b}_{nm}\Omega^z_{mn}\delta'_0
    \Big].
\end{split}
\end{equation}
Here $\delta_0=\delta(\omega-\omega_{mn})$,
$\delta'_0=\dd\delta(\omega-\omega_{mn})/\dd\omega$, and
$f^z_{nm}$ is the magnetic correction to the occupation difference. For a fixed chemical potential,
\begin{equation}
    f^z_{nm}
    =
    f'(E_n)\mathcal{M}^z_{nn}
    -
    f'(E_m)\mathcal{M}^z_{mm}.
\end{equation}
In an insulating transition with completely occupied valence bands, this occupation correction vanishes. 
The three remaining terms describe, respectively, the correction to the optical transition strength, the correction to the shift vector, and the displacement of the resonance energy. 
For an equally occupied pair of time-reversal-related valleys, the linear correction to the total shift current cancels, so the valley-summed response is even in $B_z$ to leading order.

The treatment of strong perpendicular mangetic fields requires a thorough reformulation of the model and current response on the Landau-level basis. We give a sketch of this theoretical framework in Sec.~S3 of the Supplemental Material.
In actual numerical evaluation, the strong magnetic field will be handled with magnetic supercell in a full lattice model using Peierls substitution.

%

\subsection{Full lattice models}\label{sec: model}

The analytical discussion above establishes the basic structure of the magnetic response. At zero field, the shift current is controlled by the interband connections and shift vectors of the low-energy bands. A weak in-plane field enters as a layer-dependent momentum shift, while a weak perpendicular field requires a gauge-covariant Peierls expansion in momentum space. In both cases, after summing the time-reversal-related valleys, the leading linear correction to the shift current cancels and the valley-summed response is even in the magnetic field. The full lattice model introduced below is then used to include the complete hopping structure of AB-BG and to evaluate the nonlinear current tensors numerically beyond the minimal low-energy treatment.

\begin{figure}[!htbp]
    \centering
    \includegraphics[width=0.45\textwidth]{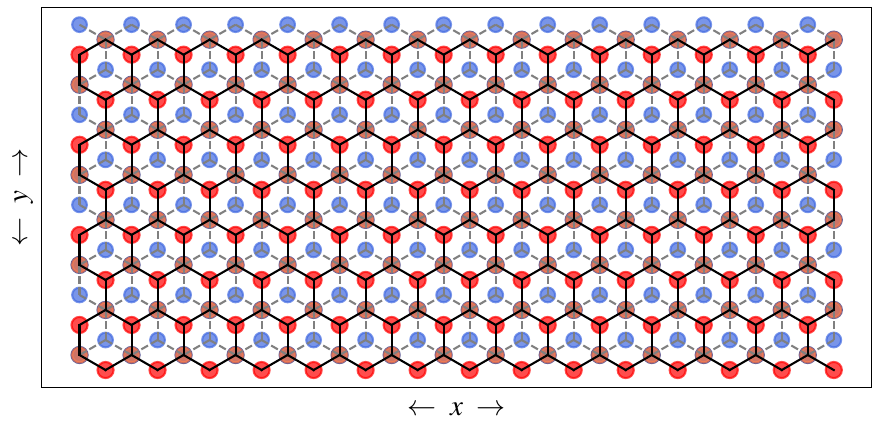}  
    \caption{Lattice structure of AB-stacked bilayer graphene. 
    Solid lines show the connection between sublattices of the \emph{top} layer, while gray dashed lines the \emph{bottom} layer. The zigzag edges of graphene flakes are aligned along the $x$-direction, while the armchair edges are along the $y$ direction. Orange dots mark the position of both $A$-sublattice of \emph{top} layer and $B$-sublattice of \emph{bottom} layer on the $xy$-plane. 
    Red dots represent the $B$-sublattice of \emph{top} layer. Light blue dots signifies the $A$-sublattice of the \emph{bottom} layer.}
    \label{fig:bg structure}
\end{figure}

For later numerical evaluation of nonlinear current responses, we will employ the full lattice model of AB-stacked bilayer graphene:
\begin{equation}
    \label{eq:abbg ham}
    \hat{H}_{\mathrm{AB-BG}}  = \sum_{\bR\alpha l, \bR' \alpha' l'} t^{\alpha l}_{\alpha' l'}(\bR+\mathbf d_{\alpha l} - \bR'-\mathbf d_{\alpha' l'}) c^\dag_{\bR \alpha l} c_{\bR' \alpha l'} ,
\end{equation}
where $l$ and $l'$ are the layer indices. $\alpha, \alpha' \in \mathrm{\{A, B\}}$ denote the sublattice indices. $\bR$ and $\bR'$ are the Bravais lattice vectors of the honeycomb lattice, with 
\begin{equation}
    \bR(n_1, n_2) = n_1 \mathbf a_1 + n_2 \mathbf a_2, ~ n_1, n_2 \in \mathbb{Z},
\end{equation}
where $\mathbf a_{1/2} =  [\pm 1/2, \sqrt{3}/2]^T \times 2.48 ~\text{\AA}$, are the base vectors of the Bravais lattice of graphene single layer. 
$\mathbf d_{\alpha l}$ indicate the relative coordinate of sublattice $\alpha$ in layer $l$ within each unit cell. Its values are enlisted here: 
$\mathbf d_\mathrm{A, top} = \mathbf d_{\mathrm{B, bottom}}= 0$, $\mathbf d_\mathrm{B, top} = 0.143 \hat{\mathbf y}~\mathrm{nm}$, and $\mathbf d_\mathrm{A, bottom} = -0.143 \hat{\mathbf y}~\mathrm{nm}$.
The hopping term is given by 
\begin{equation}
\begin{split}
    & t^{\alpha l}_{\alpha' l'}(\bR+\mathbf d_{\alpha l} - \bR'-\mathbf d_{\alpha' l'}) = \\
    & \sum_{j=1}^3 \gamma'_0 \delta_{l l'} \bar{\delta}_{\alpha \alpha'} \delta_{\bR + \mathbf d_{\alpha l} - \bR' - \mathbf d_{\alpha' l'}, \pm \mathbf d_j}  + \text{interlayer~terms}\\
\end{split}
\end{equation}
where $\bar{\delta}_{\alpha  \alpha'} = 1 - \delta_{\alpha \alpha'}$. The same convention is used for $\bar{\delta}_{l l'}$. We also define $\mathbf d_1 = d[0, 1]^T $, $\mathbf d_2 = d[-\sqrt(3)/2, -1/2]^T$ and $\mathbf d_3 = d[-\sqrt(3)/2, 1/2]^T$, where $d = 0.143~\text{nm}$ is the C-C bond length in graphene single layer. The choice of $\mathbf d_j$ vectors results in that the zig-zag edges of the graphene layer are aligned along the $x$-direction while the armchair edge is aligned along the $y$-direction.
The interlayer hopping terms are enumerated below:
\begin{eqnarray*}
    & t^{\mathrm{A,top}}_{\mathrm{A,bottom}}(\bR - \bR' + \mathbf d_1) = \gamma'_4 \sum_{j=1}^{3} \delta_{\bR + \mathbf a_{j-1}, \bR'};\\
    & t^{\mathrm{A,top}}_{\mathrm{B,bottom}}(\bR - \bR') = \gamma'_1 
    \delta_{\bR, \bR'}; \\
    & t^{\mathrm{B,top}}_{\mathrm{A,bottom}}(\bR + \mathbf d_1 - \bR') = \gamma'_3 \sum_{j=1}^{3} \delta_{\bR + \mathbf a_{j}, \bR'} ;\\
    & t^{\mathrm{B,top}}_{\mathrm{B,bottom}}(\bR + \mathbf d_1 - \bR') = \gamma'_4 \sum_{j=1}^{3} \sum_{j=1}^{3} \delta_{\bR + \mathbf a_{j-1}, \bR'} ,
\end{eqnarray*}
where we have introduced $\mathbf a_0 = \bm 0$ and $\mathbf a_3 = \mathbf a_1 + \mathbf a_2$.
The hopping amplitudes are explicitly given with $\gamma'_0 = -3.16~\mathrm{eV}$, $\gamma'_1 = -0.381~\mathrm{eV}$, $\gamma'_3 = -0.38~\mathrm{eV}$ and $\gamma'_4 = 0.14~\mathrm{eV}$, coherent with Ref~\cite{zheng2023gate}.
In Fig.\ref{fig:bg structure} we illustrate the bilayer graphene structure defined by these parameters. 

In the absence of the magnetic field, the AB-BG preserves 3-fold rotational
symmetry. The shift-current tensor transforms as
\begin{equation}
    \sigma^{abc} = R^{-1}_{aa'} \sigma^{a'b'c'} R_{b'b} R_{c'c},
\end{equation}
where 
\begin{equation*}
    R = \begin{bmatrix}
        -1/2 & -\sqrt{3}/2 \\
        \sqrt{3}/2 & -1/2
    \end{bmatrix}.
\end{equation*} 
The 3-fold rotational symmetry therefore imposes
\begin{eqnarray}
    \label{eq:independent comps1}
    \sigma^{xxx} = -\sigma^{yxy} = -\sigma^{yyx} = -\sigma^{xyy}, \\
    \label{eq:independent comps2}
    \sigma^{yyy} = -\sigma^{xxy} = -\sigma^{xyx} = -\sigma^{yxx}.
\end{eqnarray}
The application of the in-plane magnetic field breaks the $C_3$ rotational
symmetry and perturbs the equalities shown in
Eqs.~\eqref{eq:independent comps1} and~\eqref{eq:independent comps2}.

To include the parallel inplane magnetic field, we employ the Peierls substitution on the hopping terms:
\begin{equation}
\begin{split}
        &t^{\alpha l}_{\alpha' l'}(\bR+\mathbf d_{\alpha l} - \bR'-\mathbf d_{\alpha' l'})\\ 
        &\rightarrow t^{\alpha l}_{\alpha' l'}(\bR+\mathbf d_{\alpha l} - \bR'-\mathbf d_{\alpha' l'}) e^{-i \frac{e}{\hbar}\int_{\bR'+\mathbf d_{\alpha' l'}}^{\bR + \mathbf d_{\alpha l}} \bm A \cdot \dd \bm r} .
\end{split}
\end{equation}
When uniform magnetic field is parallel to the graphene planes, the vector potential is given by 
\begin{equation}
    \bm A(z) = z \bm B_{\parallel} \times \hat{\bm z}.
\end{equation}
With this convention, the vector potential is only a function of the $z$-coordinates, and is therefore constant within each layer. Performing the Fourier transform 
\[
c_{\bR \alpha l} = \frac{1}{\sqrt{N}}\sum_{\bk} e^{i\bk\cdot(\bR + \mathbf{d}_{\alpha l})} c_{\bk \alpha l} ,
\]
the Hamiltonian matrix is block diagonal in $\bk$ indices, with each diagonal block being 
\begin{equation}
    \label{eq:AB-BG}
    \begin{split}
    & H_{\mathrm{AB}}(\bk, \bm B) =\\
    & \begin{bmatrix}
        \Delta + \frac{\Delta'}{2} & \gamma'_0 g(-\bk + \bk_0) & \gamma'_4 g(\bk) & \gamma'_1 \\
        \gamma'_0 g(\bk-\bk_0) & \Delta - \frac{\Delta'}{2} & \gamma'_3 g(-\bk) & \gamma'_4 g(\bk) \\
        \gamma'_4 g(-\bk) & \gamma'_3 g(\bk) & -\Delta - \frac{\Delta'}{2} &  \gamma'_0 g(-\bk - \bk_0) \\
        \gamma'_1 & \gamma'_4 g(-\bk) &  \gamma'_0 g(\bk + \bk_0)& -\Delta + \frac{\Delta'}{2}
    \end{bmatrix} , 
    \end{split} 
\end{equation}
where 
\begin{equation}
    \bk_0 = \frac{e}{\hbar} \frac{L_z}{2} \bm A
\end{equation}
is the effective momentum shift on the \emph{top} layer with 
$L_z = 3.35~\text{\AA}$ being the vertical distance between the two graphene sheets. 
In Eq.~\eqref{eq:AB-BG}, we have chosen $z=0$ to be exactly in the middle between the two layers.
With this choice, the interlayer hopping terms acquire \emph{zero} net phase upon the Peierls substitution. 
Throughout this work we use a layer offset $\Delta=0.4~\mathrm{eV}$, which
models the vertical static field, breaks inversion, and opens a gap; pristine
AB-BG is otherwise inversion symmetric and its bulk electric-dipole shift
current vanishes. A smaller sublattice offset
$\Delta'=0.022~\mathrm{eV}$ models the hBN-induced sublattice inequivalence.
Although secondary once $\Delta\neq0$, it is retained as a realistic fixed
perturbation. All field-induced changes are measured relative to zero field
at the same offsets, thereby isolating the magnetic reconstruction. Throughout
this paper, the temperature is set to $0$ K. 
With the ingredients listed above, we can perform a scan of the magnetic field within arbitrary range. 

For AB-BG in perpendicular magnetic field, a direct bulk
tight-binding calculation requires a magnetic supercell whose area grows as $1/|B|$. 
Therefore, The weak field in the bulk 2D model will be instead treated perturbatively using the gauge-covariant Peierls expansion technique as explained in Sec.~\ref{subsec:perpendicular-weak-field} to circumvent the difficulty with extremely large supercells. 
The calculation at strong field will be achieved using the magnetic supercell with appropriate choices of field strength.
As an essential complement to the calculations in the bulk, the ribbon models allow for calculations at arbitrary strength of the perpendicular magnetic field. 
More impportantly, the effect of the edge states can only manifest itself with the ribbon models.


\section{Results and discussion}
\label{sec: results}

\subsection{AB-bilayer graphene in in-plane magnetic field}

\begin{figure}[!htbp]
    \centering
    \includegraphics[width=0.45\textwidth]{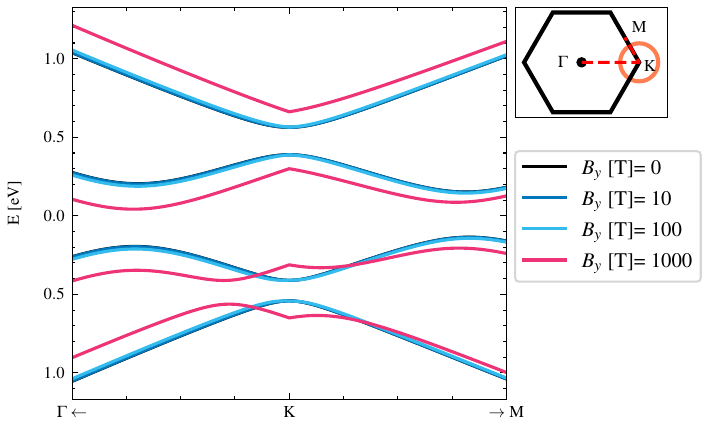}
    \caption{Band structure of AB-BG near the K-valley under different magnetic field along $y$ direction with intensity ranging from $0$ to 1000 Tesla. The band structures are \emph{not} shown for the whole route $\mathrm{\Gamma-K-M}$. 
    Only the part inside in the orange circle is presented to concentrate on the band structures near the Fermi level (as we have keep the chemical potential $\mu = 0$). No substantial change can be found in the band structures unless the parallel magnetic field is unrealistically strong.}
    \label{fig:bg band Bpara}
\end{figure}

In Fig.~\ref{fig:bg band Bpara} we show that under the application of a magnetic field the electronic band structure remains essentially unchanged unless the field strength is pushed to unrealistically large values (on the order of $10^3$~T). Despite the barely visible influence on the band dispersions themselves, the magnetic field nevertheless reduces spatial symmtry of the system and breaks TRS. 
This symmetry reduction can have nontrivial consequences for the nonlinear current response, which is the central focus of our investigation.

\begin{figure}[!htbp]
    \centering
    \includegraphics[width=0.95\linewidth]{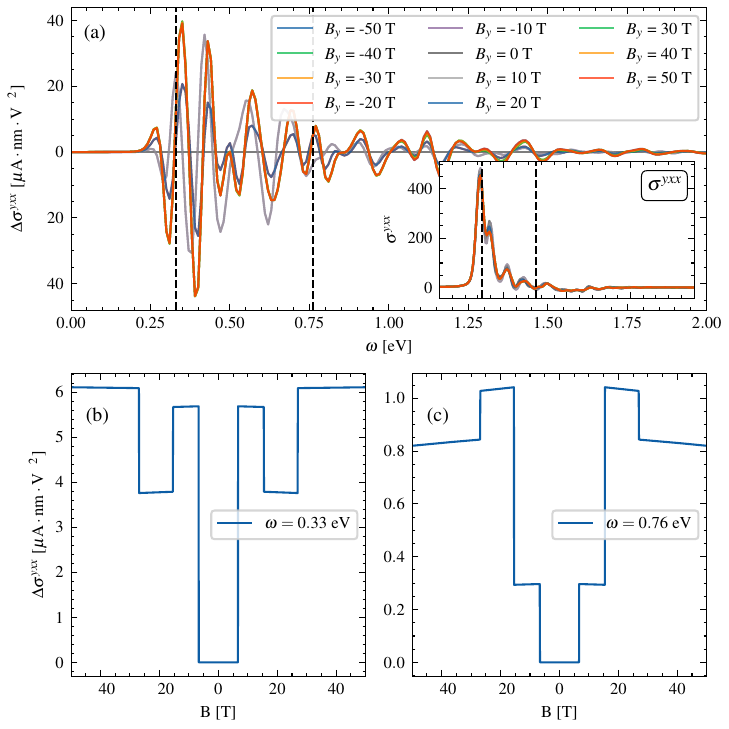}
    \caption{(a) The variation of $\sigma^{yxx}$ with respect to zero magnetic field. In the inset plot of (a) is shown the complete curve of $\sigma^{yxx}$ under different magnetic field. Among the $2000$ evenly distributed field strength values between $\pm 50~\mathrm{T}$ sampled in our calculations, we plot the data every $10~\mathrm{T}$ for better readability.
    The vertical dashed lines indicate the photon energies at which we track the variation of $\sigma^{yxx}$ in (b) and (c), respectively. (b) The variation of $\sigma^{yxx}$ with photon energy fixed at $\omega = 0.33~\mathrm{eV}$. (c) The variation of $\sigma^{yxx}$ with photon energy fixed at $\omega = 0.76~\mathrm{eV}$.}
    \label{fig:sigma yxx BG}
\end{figure}

In the absence of magnetic field, the most remarkable shift current response is $\sigma^{yxx}$.
We focus on its variation under different strengths of the magnetic field.
We first align the magnetic field along the $y$-direction and vary its strength from $-50$ to $50$ Tesla, with a step of $0.05~\mathrm{T}$. We show the evolution of $\sigma^{yxx}$ under different strengths of the magnetic field in Fig.~\ref{fig:sigma yxx BG}. 
The shift current response is an \emph{even} function of the magnetic field, as suggested from the theoretical analysis. 
But here, the detail scan of the inplane magnetic field does not suggest a parabolic relation between shift current and the magnetic field, as shown in Fig.~\ref{fig:sigma yxx BG}(b) and (c).
where we track the variation of the largest peak located at $\omega = 0.33~\mathrm{eV}$ and $0.76~\mathrm{eV}$, respectively. 
The in-plane magnetic field redistributes the shift-current spectrum only modestly. The change $\Delta\sigma^{yxx}$ is symmetric under $B_y \rightarrow -B_y$, consistent with the even-field character of the shift current, but the fixed-frequency traces do not follow a simple universal parabolic law. Instead, their detailed shape depends on photon energy and appears step-like, reflecting the motion of sharp spectral features through the finite energy bin used in the calculation.

\subsection{AB-BG in weak perpendicular magnetic field}

\begin{figure}[!htbp]
\centering
\includegraphics[width=0.9\linewidth]{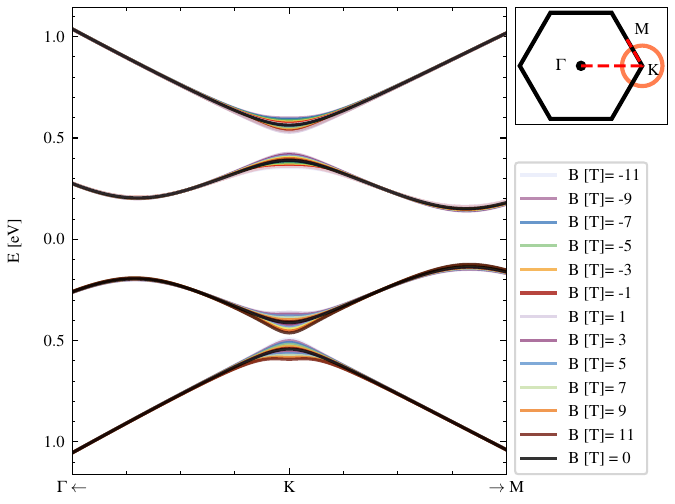}
\caption{Band structure of AB-BG in weak perpendicular magnetic fields treated perturbatively. 
Similar to Fig.~\ref{fig:bg band Bpara}, the band structure near $K$ point is shown along the path indicated on top right of the figure. Band perturbation is visible even the field is weak, and is mostly visible near the lifted Dirac points where the Berry curvature is concentrated. The deformation of band happens mostly at the lift of the Dirac points, where the Berry curvature is stongly perturbed by the external field.}  
\label{fig: BG weak perp B} 
\end{figure}

The weak perpendicular magnetic field is treated using the gauge-covariant Bloch expansion as explained in Sec.~\ref{subsec:perpendicular-weak-field}. 
Using Eq.~\eqref{eq: ham weak perp B} to take into account the perpendicular magnetic field to linear order and diagonalizing the Hamliltonian, we obtain the band structures for various field intensities, as presented in Fig.~\ref{fig: BG weak perp B}.
We remark that the band deformation happens the most strongly where the Berry curvature is the most concentrated, i.e. the vicinity of $K$ and $K'$ points. 

\begin{figure}[!htbp]
\centering
\includegraphics[width=\linewidth]{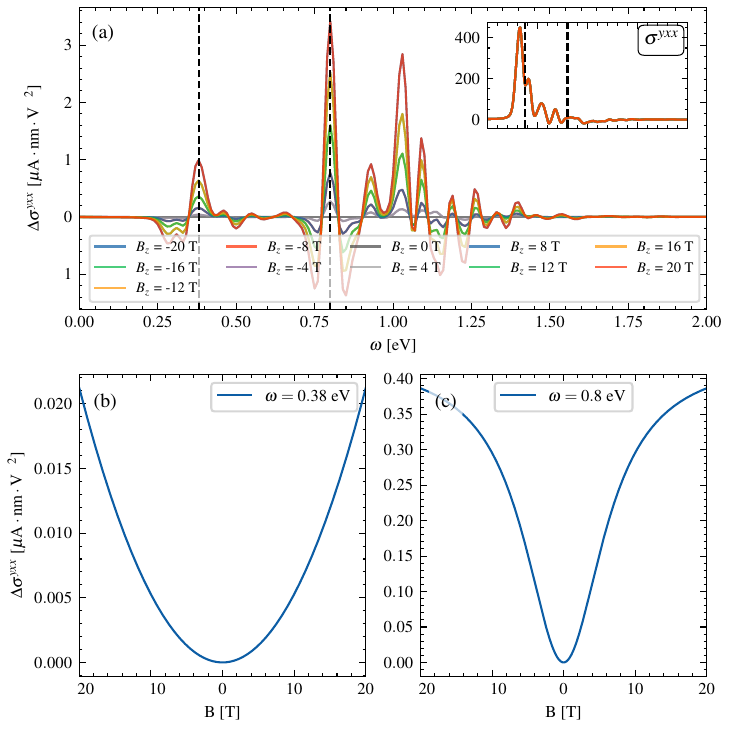}
\caption{(a) The variation of $\sigma^{yxx}$ with respect to zero magnetic field. In the inset plot of (a) we also show the complete curve of $\sigma^{yxx}$ under different magnetic field. 
A scan of the magnetic field was performed for 800 point ranging from $B_z = -20~\mathrm{T}$ to  $B_z = 20~\mathrm{T}$, with a step size of $0.05~\mathrm{T}$.
We plot the data every $5~\mathrm{T}$ for better readability.
The vertical dashed lines indicate the photon energies at which we track the variation of $\sigma^{yxx}$ in (b) and (c), respectively. (b) The variation of $\sigma^{yxx}$ with photon energy fixed at $\omega = 0.38~\mathrm{eV}$. (c) The variation of $\sigma^{yxx}$ with photon energy fixed at $\omega = 0.8~\mathrm{eV}$.}
\label{fig: sc yxx BG weak perp B}
\end{figure}

\begin{figure}[!htbp]
    \centering
    \includegraphics[width=0.96\linewidth]{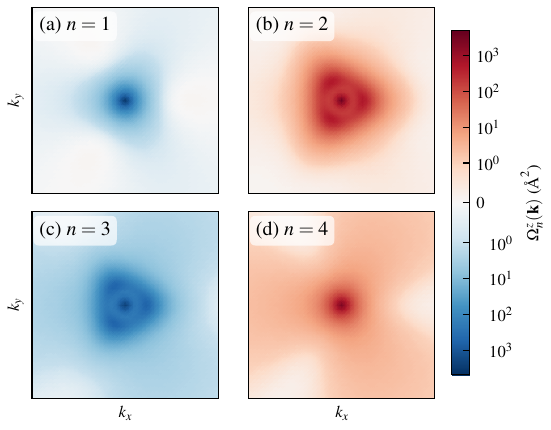}
    \caption{Berry-curvature distribution near the $K$ point of AB-stacked bilayer graphene at zero magnetic field. Panels (a)--(d) show $\Omega_n^z(\bk)$ for bands $n=1,\ldots,4$, respectively, ordered from the lowest to the highest energy. Red and blue colors denote positive and negative Berry curvature, respectively. All panels use the same symmetric logarithmic color scale, with the Berry curvature reported in $\text{\AA}^2$. 
    The Berry curvature is sharp and peaked near the lifted Dirac points with weaker surrounding distribution. 
    The hottest regions of Berry curvature coincide with the onset of the largest band deformation under weak perpendicular magnetic field.}
    \label{fig:berry curvature}
\end{figure}

In Fig.~\ref{fig:berry curvature}, we report the Berry curvature distribution on each Band near the $K$ point. It is straightforward to see that the strongest band deformation caused by the perpendicular magnetic field happens where the Berry curvature is the most concentrated. For this four band model without band crossing, the Berry curvature is conveniently evaluated using Eq.~(S1) of the Supplemental Material.

Figure~\ref{fig: sc yxx BG weak perp B} shows the change of the bulk shift-current component $\sigma^{yxx}$ under a weak perpendicular magnetic field. 
Despite strong concentration of Berry curvature near the $K$ and $K'$ points, the strongest shift current response happens elsewhere, from the transition between the band edges.
Compared with the zero-field spectrum, the field mainly redistributes spectral weight around the existing optical resonances, while the overall magnitude of the correction remains small. As required by the time-reversal-related valley symmetry discussed above, $\Delta\sigma^{yxx}$ is symmetric under $B_z\rightarrow -B_z$. The fixed-frequency traces in Figs.~\ref{fig: sc yxx BG weak perp B}(b) and (c) show that, close to $B_z=0$, the leading variation is quadratic in $B_z$, consistent with the absence of a linear correction to the valley-summed shift current. 
Away from the immediate weak-field limit, however, the detailed profile of the $\sigma^{a;bb}$--$B_z$ relation depends also on the photon frequency because the perpendicular field shifts and reshapes the resonant optical transitions, and more importantly, the local Berry curvature.

\subsection{AB-BG in strong perpendicular magnetic field}

In the strong perpendicular-field regime, the magnetic flux through a primitive cell is no longer perturbative. We therefore treat the bulk lattice model non-perturbatively by using a magnetic supercell. The flux through one primitive graphene cell is chosen to be a rational fraction of the flux quantum, $\Phi=B_z S_0=(p/q)\Phi_0$, where $S_0$ is the primitive-cell area and $\Phi_0=h/e$ is the magnetic flux quantum. The real-space unit cell is then enlarged to contain $q$ primitive cells, so that the total flux through the magnetic supercell is the integer value $p\Phi_0$ and magnetic translation symmetry is recovered. The field enters the tight-binding Hamiltonian through Peierls phases on the hoppings, with the phase accumulated around each primitive plaquette equal to $2\pi p/q$.

\begin{figure}[!htbp]
\centering
\includegraphics[width=0.6\linewidth]{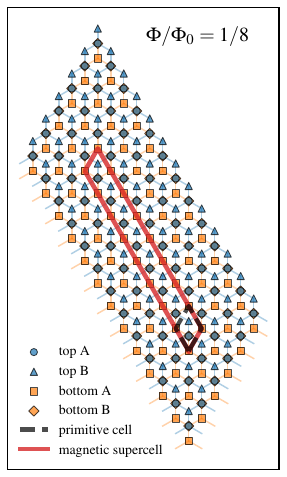}
\caption{Sketch of the magnetic supercell for a per-plaquette magnetic flux $\Phi = \Phi_0 / 8$. Note that this amounts to an extreme magnetic field strength of $\sim 10^4$ T.}
\label{fig:magnetic supercell}
\end{figure}

The construction is illustrated in Fig.~\ref{fig:magnetic supercell} for $p/q=1/8$, corresponding to eight primitive cells in the magnetic supercell. This approach preserves a Bloch description and allows the shift current to be computed with the same band-based formalism used at zero field, without reformulating the response directly in a Landau-level basis.

\begin{figure}[!htbp]
\centering
\includegraphics[width=0.95\linewidth]{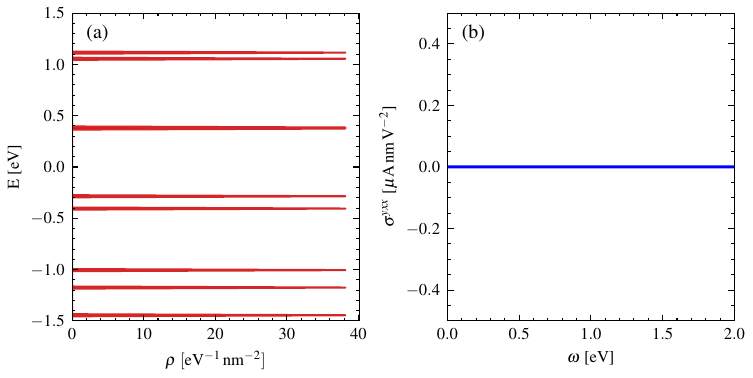}
\caption{Using a supercell with $q = 50$, which corresponds to $B_z \sim 1580$ T, we show in (a) the density of states of the highly quantized Landau levels and in (b) the nearly complete suppression of the shift current under strong Landau quantization.}
\label{fig: sc in strong B}
\end{figure}

For the explicit calculation shown in Fig.~\ref{fig: sc in strong B}, we use $p/q=1/50$, corresponding to $B_z\simeq 1580~\mathrm{T}$. At this field strength the spectrum is reorganized into nearly flat Landau-level-like bands, producing sharp structures in the density of states. A large density of states might naively suggest a large joint density of states (JDOS) and therefore an enhanced optical response. The computed shift current shows the opposite trend: $\sigma^{yxx}$ is almost completely quenched over the optical-energy range considered.

This result highlights that the bulk shift current cannot be inferred from the (joint) density of states alone. Once Landau quantization dominates, the optical matrix elements and shift-vector factors entering the shift-current response are strongly reduced or mutually cancel, so the high degeneracy of the Landau-level spectrum does not translate into an enhanced bulk photovoltaic response.

\subsection{AB-BG nanoribbons in vertical magnetic fields}

In the following we explore how the shift current response in AB-BG nanoribbons can be altered by the perpendicular magnetic field. 
The zigzag and armchair ribbons are constructed from the structure shown in Fig.~\ref{fig:bg structure} by making the system finite along the $y$ and $x$ directions, respectively. 
The nanoribbons are of particular interest compared to the bulk model, because the nanoribbons may host particular electronic properties and magnetic tunabilities in comparison with the 2D bulk model due to the finite width and the explicit edges.
Depending on how the nanoribbon is cut out from the 2D flake, graphene nanoribbons can be in principle classifed as zigzag and armchair ribbons, according to the atom lines near the edge of the ribbon. 
It has been established that in the absence of external magnetic field, edge states naturally exist in zigzag-edge graphene nanoribbon while the armchair ribbon only possess dispersive states.~\cite{castro2009electronic} 
Before performing any detailed computation, we can go through some useful speculations about how magnetic field effects can differ in the nanoribbons compared to the 2D bulk system.
Magnetic field, especially strong magnetic fields form Landau levels in the bulk and suppresses shift current. 
In the zigzag nanoribbons, the originally localized edge states may acquire considerable space extent under strong perpendicular magnetic field and result in non-zero dipole matrix elements.
In the meantime, while the Landau-level states are in general localized, some may also acquire finite dispersion as the guiding center approaches the edges and therefore contribute to the optical response. They can be noted as Landau-level edge states.
These effects concerning the two types of edges states are completely invisible in the bulk model. 
Therefore, the ribbon models should be regarded as an important complement of the prevous studied 2D models, rather than their trivial simplifications.
On the other hand, the nanoribbon models do offer a technical convenience to study the effects of perpendicular magnetic field. 
A continous scan from weak to extremely strong magnetic fields can be performed seamlessly for arbitary values of field magnitudes, without constructing large magnetic supercells or diving into the Landau-level formulations (see Sec.~S3 of the Supplemental Material). 

In the following calculations, the ribbon models consist of $20$ zigzag or armchair atomic lines in each graphene layer. With this setup. the width of the zigzag ribbon is fixed to be $41.9~\text{\AA}$ while the armchair ribbon is $23.4~\text{\AA}$ wide.

\subsubsection{Zigzag AB-BG nanoribbon}

\begin{figure}[!htbp]
    \centering
    \includegraphics[width=0.45\textwidth]{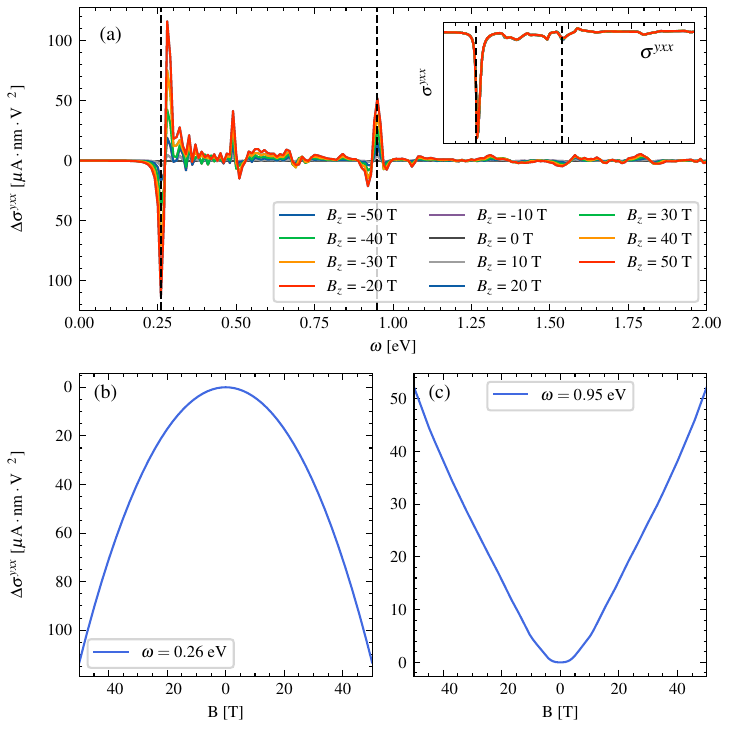}
    \caption{(a) The main plot shows the variation of $\sigma^{yxx}$ with respect to the zero-field profile under different strengths of the perpendicular mangetic field in the AB-stacked zigzag nanoribbon. 
    The complete profiles of $\sigma^{yxx}$ are presented in the inset. 
    Among the $2000$ evenly sampled data points, we plot the shift current curves every $10~\mathrm{T}$ for better readability. 
    The vertical dashed lines indicate the photon energies at which we track the variation of $\sigma^{yxx}$ in (b) and (c), respectively. 
    (b) The variation of $\sigma^{yxx}$ with photon energy fixed at $\omega = 0.27~\text{eV}$. (c) The variation of $\sigma^{yxx}$ with photon energy fixed at $\omega = 0.95~\text{eV}$.}
    \label{fig:ribbon sc}
\end{figure}
The band structures and shift current responses computed for the zigzag nanoribbon at $B_z = 0$ and $B_z = 1000$ T can be found in Secs.~S4 and S5 of the Supplemental Material, respectively.
It is noteworthy that even under strong $B_z$, the nanoribon with finite width retains certain band dispersion, and the shift current response is not entirely suppressed. 
Such effects due to the finite width are not visible in the 2D bulk model studies previously.

In Fig.~\ref{fig:ribbon sc} we show the shift current conductivity component $\sigma^{yxx}$ under different $B_z$ strengths ranging from $-50$ to $50$ T. 
Examining the resonant frequencies and compare them to the gap sizes shown in Fig.~S1 of the Supplemental Material, one realizes that the edge states (characaterized by the flattened parts in the band structure) has exactly zero contribution to the shift current response, despite their tempting JDOS profile indicated in Fig.~S2 of the Supplemental Material.

We trace the variation of $\sigma^{yxx}$ at two fixed photon frequencies, $\omega = 0.26$ eV and $\omega = 0.95$ eV in Figs.~\ref{fig:ribbon sc} (b) and (c) respectively.
We observe again the variation of the shift current conductivity as an even function of the magnetic field, as it should. 
But the detailed relation between the shift current and $B_z$ differs by light frequency. In the close vicinity of $B_z = 0$, the smooth curves can be approximated by a parabolic relation. 
But going farther beyond, more various funcitonal relation appears and depend strongly on the photon frequency.

\begin{figure}[!htbp]
    \centering
    \includegraphics[width=0.45\textwidth]{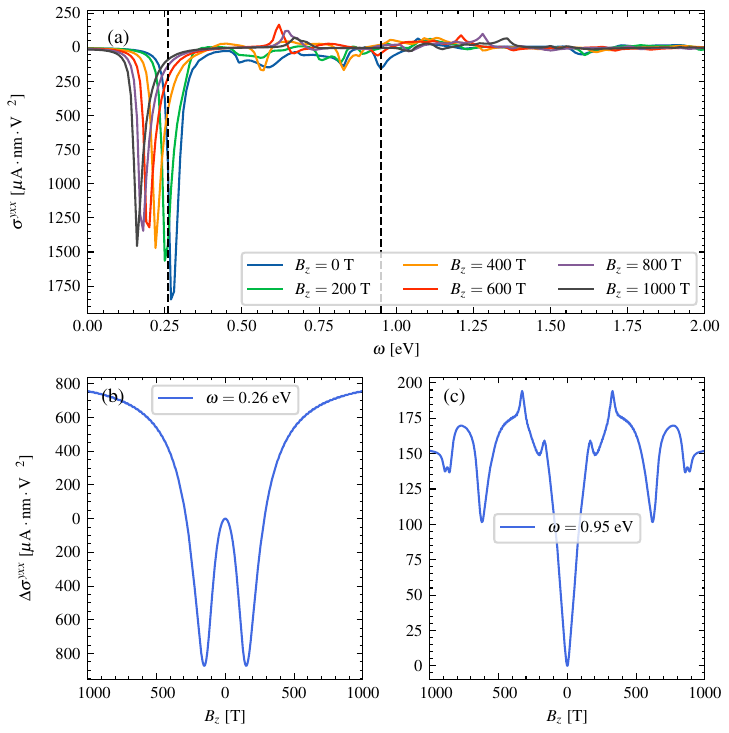}
    \caption{Evolution of the zigzag-nanoribbon shift-current response over the full perpendicular-field range. (a) Shift-current spectra $\sigma^{yxx}(\omega)$ at $B_z=0$, $200$, $400$, $600$, $800$, and $1000~\mathrm{T}$. The vertical dashed lines mark the photon energies selected in the lower panels. (b) and (c) Field-induced changes $\Delta\sigma^{yxx}(B_z)=\sigma^{yxx}(B_z)-\sigma^{yxx}(0)$ from $-1000$ to $1000~\mathrm{T}$ at $\omega=0.26~\mathrm{eV}$ and $0.95~\mathrm{eV}$, respectively.}
    \label{fig:ribbon sc full B}
\end{figure}

To gain a complete view of the evolution of the shift current from weak to strong $B_z$, we extend the scan to the range $-1000\leqslant B_z\leqslant 1000~\mathrm{T}$. The dense sampling used between $-50$ and $50~\mathrm{T}$ is retained, while a step size of $1~\mathrm{T}$ is adopted outside this interval. 
The results are summarized in Fig.~\ref{fig:ribbon sc full B}

Figure~\ref{fig:ribbon sc full B} shows that the even-field relation observed over the entire interval with dense the narrow-range scan preserved.
However, the approximately quadratic behavior found close to $B_z=0$ in Fig.~\ref{fig:ribbon sc} is only a local weak-field property and cannot be extrapolated to the strong-field regime. 
As shown by the perturbative decomposition in Sec.~\ref{subsec:perpendicular-weak-field}, the magnetic field modifies several ingredients of the shift-current integrand simultaneously: the orbital magnetic correction shifts the interband transition energies and therefore displaces the optical resonances; the explicit correction to the velocity operator and the magnetic mixing of the wave functions change the optical transition strength; and the field-induced changes of the intraband Berry connections and of the phase of the interband connection modify the shift vector. 

In the strong-field regime, higher-order magnetic mixing and the progressive formation of Landau-level-like states become important; in a finite ribbon, the accompanying reconstruction and spatial broadening of the edge states further alter the interband matrix elements and shift vectors. The resulting competition explains the non-monotonic and strongly frequency-dependent evolution, together with the displacement and reconstruction of the optical resonances visible in Fig.~\ref{fig:ribbon sc full B}(a). Even at $B_z=1000~\mathrm{T}$, the nanoribbon retains a finite shift-current spectrum, in contrast to the nearly complete suppression obtained for the strongly quantized two-dimensional bulk system.

In particular, the prominent low-energy peak of $\sigma^{yxx}$ survives as $B_z$ is increased from zero to the strong-field regime. To identify the states underlying this resonance, we search the two middle bands at $B_z=1000~\mathrm{T}$ for the smallest direct gap and examine the corresponding wave functions.
\begin{figure}[!htbp]
    \centering
    \includegraphics[width=0.9\linewidth]{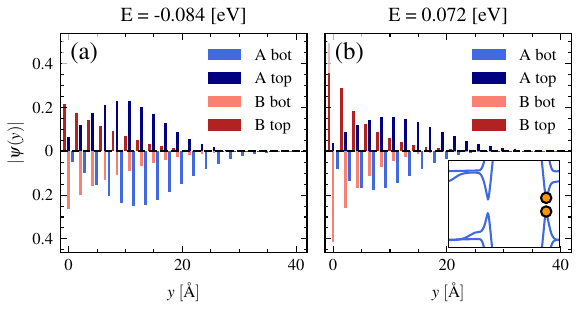}
    \caption{Wave-function amplitudes of the two middle-band states forming the minimum direct gap of the $N_{\mathrm{zig}}=20$ AB-BG zigzag nanoribbon at $B_z=1000~\mathrm{T}$. Panels (a) and (b) show, respectively, the valence state at $E_v=-0.084~\mathrm{eV}$ and the conduction state at $E_c=0.072~\mathrm{eV}$, evaluated at one of the minimum band gaps (marked by the two dots in the band structure plot in the inset of (b)). 
    Their energy separation is $E_c-E_v=0.156~\mathrm{eV}$. 
    Blue and red bars denote amplitudes on sublattices $A$ and $B$, while light and dark shades distinguish the bottom and top layers. For visual separation, top-layer amplitudes are plotted above the horizontal axis and bottom-layer amplitudes below it; the tick labels give their absolute values.}
    \label{fig:min direct gap wfs}
\end{figure}

As shown in Fig.~\ref{fig:min direct gap wfs}, neither state is confined to a single atomic line, layer, or sublattice. Both retain appreciable weight over a common region extending from the edge into the ribbon interior, and their multicomponent distributions overlap spatially. The strong perpendicular field therefore reconstructs the edge-derived states into sufficiently extended states with non-negligible interband optical matrix elements, rather than the strongly localized and layer--sublattice-polarized states that are effectively dark in the weak-field regime. The minimum direct gap of approximately $0.156~\mathrm{eV}$ also coincides with the photon-energy range of the surviving low-energy peak in $\sigma^{yxx}$. This agreement identifies the transition between this valence--conduction pair as a principal microscopic channel for the strong-field shift-current response. The result further illustrates why Landau-level formation alone does not determine the response magnitude: the spatial structure, layer--sublattice composition, interband connection, and shift vector of the participating states remain essential.

The pair of states are reminiscent of the edges states as they show limited spatial distribution close to only one edge of the ribbon. To prove this, we explore how their shift current response varies with the ribbon width. 

\begin{figure}[!htbp]
    \centering
    \includegraphics[width=0.45\textwidth]{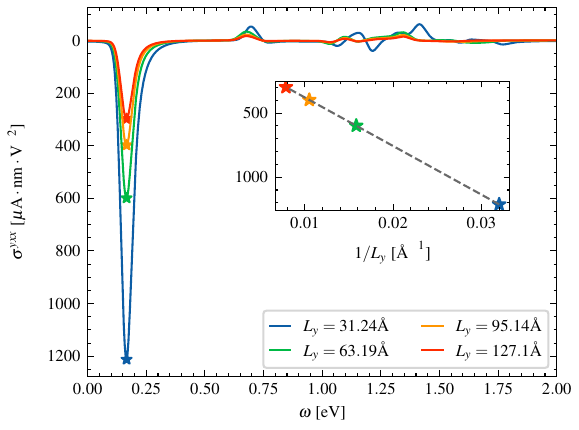}
    \caption{$\sigma^{yxx}$ with different ribbon widths. Vertical magnetic field is fixed at $1000$ T. The inset show the plot of the peak intensity as a function of the inverse of the ribbon width, demonstrating perfect linearity. The inset plot validates the dilution effect by increasing the ribbon width, and therefore supports our argument that the intense peak originates solely from the edge states.}
    \label{fig:ll dilution}
\end{figure}
The number of the edge states is fixed, and the DOS and JDOS from them will be diluted as the width of the nanoribbon grows. 
Therefore, we expect the intense peak to be \emph{inversely} proportional to the width. 
In Fig.~\ref{fig:ll dilution}, we compute and show $\sigma^{yxx}$ for different widths of the nanoribbon. In the inset we show the peak intensity as a function of the $1/L_y$ with $L_y$ denoting the width of the ribbon. A perfect linear relation is found between the peak intensity and the inverse of the ribbon width, proving that the sharp intense peak is due to the transition between edge states. 
Therefore, we have found that in the zigzag nanoribbon the edge states play exactly opposite roles under weak and strong magnetic field in the shift current response -- dark in weak field but bright in strong field.
We remark that the other weak but nonzero shift current responses at higher light frequencies also decreases as the width grows, indicating they also come from edge states from higher bands.

\begin{figure}[!htbp]
    \centering
    \includegraphics[width=0.9\linewidth]{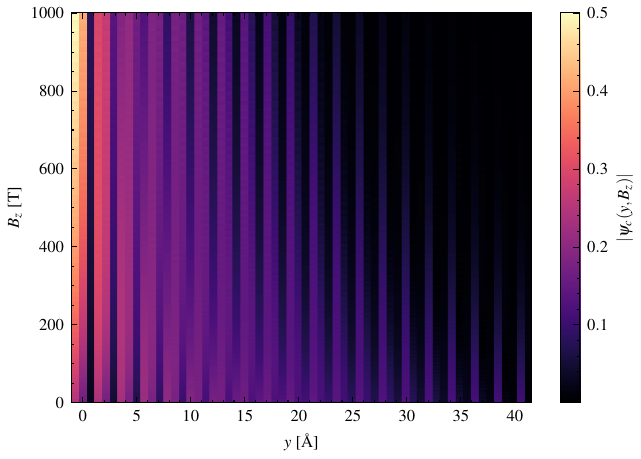}
    \caption{Evolution of the minimum-gap conduction-band wave function in the $N_{\mathrm{zig}}=20$ AB-BG zigzag nanoribbon as a function of perpendicular magnetic field. For every $B_z$ from $0$ to $1000~\mathrm{T}$ in steps of $1~\mathrm{T}$, the smallest direct gap between the two middle bands is located within the $k_x>0$ half of the one-dimensional Brillouin zone, and the normalized conduction state at that momentum is recorded. The color map shows $|\psi_c(y,B_z)|=\sqrt{\sum_{i:y_i=y}|\psi_{c,i}(B_z)|^2}$, where the sum combines the layer- and sublattice-resolved orbitals sharing the same transverse coordinate $y$.}
    \label{fig:edge state brightening}
\end{figure}

To further investigate the onset of the edge-state brightening as the $B_z$ strengthens, we track the evolution of the wave functions at the conduction band edge as a function of $B_z$, as shown in Fig.~\ref{fig:edge state brightening}.
As the field strength increases, the states at band edges responding to the shift current become more and more localized to the edge. 
Fig.~\ref{fig:edge state brightening} shows a smooth transition in the brightest states from diffusive states to localized edge states. The edge-state brightening happens roughly near $B_z = 350$ T where the wave function starts to contract to one of the edges of the ribbon.

\subsubsection{Armchair AB-BG nanoribbon}

\begin{figure}[!htbp]
    \centering
    \includegraphics[width=0.95\linewidth]{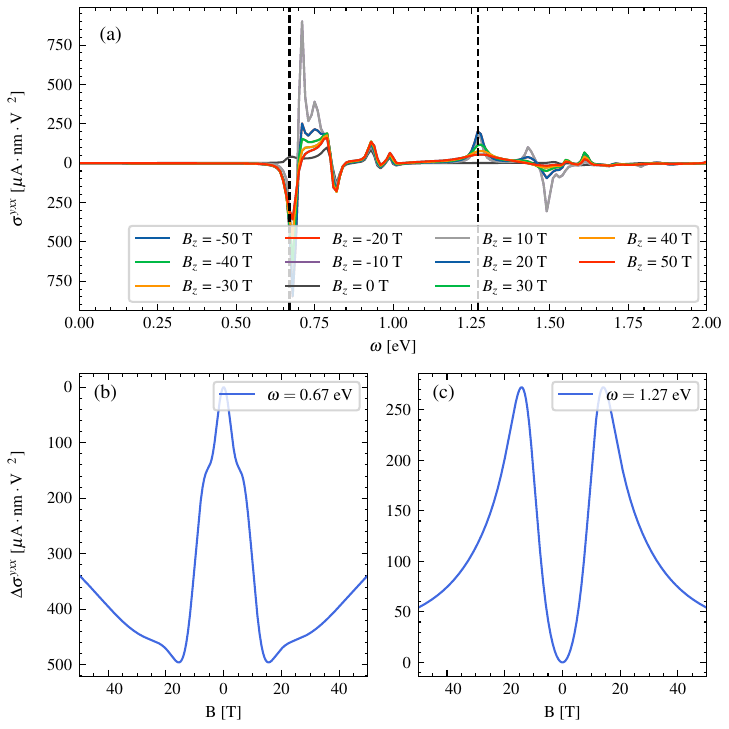}
    \caption{Weak-field evolution of the shift-current response in the $N_{\mathrm{arm}}=20$ AB-BG armchair nanoribbon. (a) Spectra of $\sigma^{yxx}(\omega)$ from $B_z=-50$ to $50~\mathrm{T}$, shown every $10~\mathrm{T}$ for clarity. The vertical dashed lines mark the photon energies used in the lower panels. (b) and (c) Field-induced changes $\Delta\sigma^{yxx}(B_z)=\sigma^{yxx}(B_z)-\sigma^{yxx}(0)$ at $\omega=0.67~\mathrm{eV}$ and $1.27~\mathrm{eV}$, respectively.}
    \label{fig:ribbon arm sc}
\end{figure}

We perform the same scan of $B_z$ from $-1000$ to $1000$ T with a step of 1 T and the grid is densified between $\pm 50$ T with a smaller step size of 0.05 T.
Figure~\ref{fig:ribbon arm sc} focuses on this densely sampled weak-field interval.

The perpendicular field substantially reconstructs the resonances between approximately $0.6$ and $1.6~\mathrm{eV}$, rather than producing a uniform enhancement or suppression of the spectrum. At the two representative photon energies, $\Delta\sigma^{yxx}$ is even under field reversal, consistent with the symmetry found above: the response at $0.67~\mathrm{eV}$ is suppressed, whereas that at $1.27~\mathrm{eV}$ is enhanced. Both traces are strongly nonmonotonic away from $B_z=0$, with extrema near $|B_z|\simeq 15~\mathrm{T}$, showing that the field shifts and reshapes nearby optical resonances.

Unlike the zigzag ribbon, the armchair ribbon has no localized edge band at zero field. The behavior in Fig.~\ref{fig:ribbon arm sc} therefore does not reflect the weak-to-strong-field brightening of a pre-existing flat edge band discussed above. Instead, it provides a complementary example in which magnetic reconstruction of dispersive ribbon states is sufficient to produce a large, frequency-selective change of the shift current while preserving the required even-field symmetry.

\begin{figure}[!htbp]
    \centering
    \includegraphics[width=0.95\linewidth]{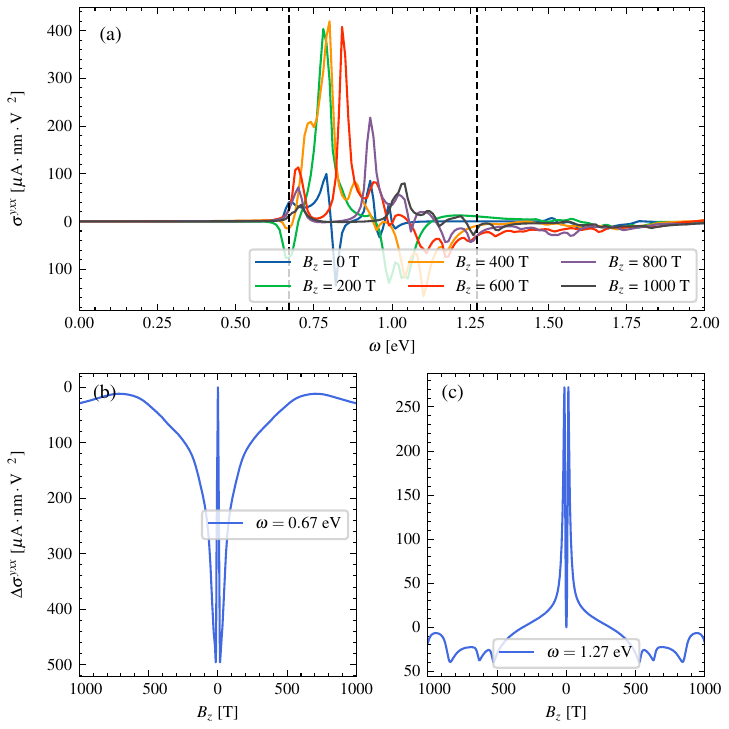}
    \caption{Evolution of the armchair-nanoribbon shift-current response over the full perpendicular-field range. (a) Shift-current spectra $\sigma^{yxx}(\omega)$ at $B_z=0$, $200$, $400$, $600$, $800$, and $1000~\mathrm{T}$. The vertical dashed lines mark the photon energies selected in the lower panels. (b) and (c) Field-induced changes $\Delta\sigma^{yxx}(B_z)=\sigma^{yxx}(B_z)-\sigma^{yxx}(0)$ from $-1000$ to $1000~\mathrm{T}$ at $\omega=0.67~\mathrm{eV}$ and $1.27~\mathrm{eV}$, respectively.}
    \label{fig:ribbon arm sc full B}
\end{figure}

Figure~\ref{fig:ribbon arm sc full B} places the weak-field behavior in the context of the complete field scan. The fixed-frequency traces remain even under $B_z\rightarrow -B_z$ throughout the interval as expected.
but their evolution is highly nonmonotonic. At $0.67~\mathrm{eV}$, the pronounced suppression near the weak-field extrema gradually weakens at larger $|B_z|$. At $1.27~\mathrm{eV}$, the initial enhancement changes sign as the field increases and is followed by additional smaller structures in the strong-field regime. Thus, neither the sign nor the magnitude of the field-induced change inferred close to $B_z=0$ can be extrapolated to high fields.

The spectra in Fig.~\ref{fig:ribbon arm sc full B}(a) show the corresponding reconstruction in frequency space. The dominant resonances shift toward higher photon energies as $B_z$ increases, while their amplitudes first grow and then decrease. A finite response nevertheless remains at $B_z=1000~\mathrm{T}$, again contrasting with the nearly complete quenching found in the strongly quantized two-dimensional bulk. Since the armchair ribbon has no pre-existing flat edge band at zero field, this surviving response should not be identified with the same edge-state-brightening mechanism established above for the zigzag ribbon.

\begin{figure}[!htbp]
    \centering
    \includegraphics[width=0.95\linewidth]{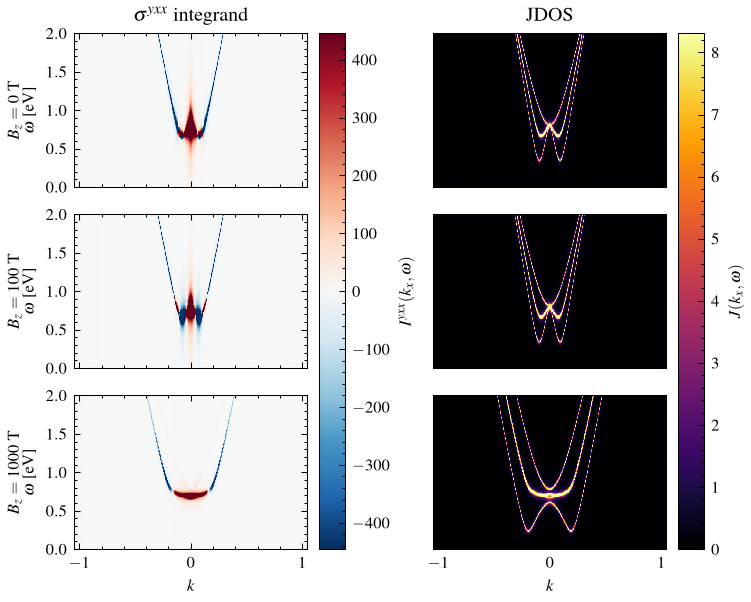}
    \caption{Energy- and momentum-resolved optical response of the $N_{\mathrm{arm}}=20$ AB-BG armchair nanoribbon. The rows correspond to $B_z=0$, $100$, and $1000~\mathrm{T}$, from top to bottom. The left column shows the signed shift-current integrand $I^{yxx}(k,\omega)$, whose integral over the one-dimensional Brillouin zone yields $\sigma^{yxx}(\omega)$; red and blue denote positive and negative contributions, respectively. The right column shows the corresponding joint density of states $\mathcal{J}(k,\omega)$. Here, $v$ ($c$) is the highest valence (lowest conduction) band, $v-1$ is the next valence band below $v$, and $c+1$ is the next conduction band above $c$. Both quantities include the four central interband transitions $(v,c)$, $(v-1,c)$, $(v,c+1)$, and $(v-1,c+1)$ and are normalized by the ribbon width. A common color scale is used within each column to facilitate comparison between magnetic fields.}
    \label{fig:intgrand and jdos}
\end{figure}

We label the highest valence band as $v$ and the lowest conduction band as $c$; $v-1$ denotes the valence band immediately below $v$, while $c+1$ denotes the conduction band immediately above $c$. Although the gap between $v$ and $c$ is only $\sim 0.3~\mathrm{eV}$, the first major peak in the shift-current spectrum appears near $0.68~\mathrm{eV}$, indicating that the direct $(v,c)$ transition contributes only weakly to the integrated response. To resolve the relevant transitions, we plot the shift-current integrand and JDOS as functions of photon energy and crystal momentum in Fig.~\ref{fig:intgrand and jdos}.

\section{Conclusion}
\label{sec: conclusion}

We have established a comprehensive description of the shift-current response of inversion-broken AB-stacked bilayer graphene from weak to strong magnetic fields. 
The approach combines the layer-dependent momentum shift generated by an in-plane field, a gauge-covariant Peierls expansion for weak perpendicular fields, a magnetic-supercell construction for the nonperturbative bulk regime, and finite-width ribbon calculations that remain applicable at arbitrary field strength. 
This framework makes it possible to follow not only the evolution of the spectrum, but also the magnetic reconstruction of the wave functions and nonlinear optical matrix elements.

In the two-dimensional bulk system, the field orientation leads to qualitatively different orbital effects. An in-plane field acts through the small separation between the graphene layers and produces opposite momentum shifts in the two layers. 
It therefore leaves the band dispersion nearly unchanged at realistic fields, while lowering the crystalline symmetry and
modestly redistributing the shift-current spectrum. 
A perpendicular field couples directly to the in-plane orbital motion. Its weak-field corrections are concentrated near Berry-curvature hot spots, whereas sufficiently strong fields reorganize the spectrum into nearly flat Landau-level-like bands. 
For both orientations, the valley-summed shift current is even under field reversal, so a smooth weak-field expansion has no linear term. 
The detailed variation nevertheless remains strongly frequency dependent because the field displaces and reshapes individual optical resonances.

One fundamental implication of our results is a concrete magnetic-field
realization of the established principle that the density of available
transitions alone is not a reliable measure of the nonlinear photovoltaic
response~\cite{sipe2000second-order,cook2017design}. A large shift current
requires resonant phase space, a finite interband connection, and a finite
shift vector at the same momentum and photon energy. 
Consequently, the Berry-curvature hot spots that mark the
largest weak-field band corrections need not coincide with the dominant shift-current channels. 
More strikingly, strong Landau quantization greatly enhances the bulk DOS and JDOS but almost completely quenches the bulk shift current because the relevant optical matrix elements and shift-vector contributions are suppressed or mutually cancel. Magnetic fields should therefore be viewed as tools for engineering wave-function geometry and optical activity, rather than as a means of increasing the number of
available transitions alone.

Boundaries qualitatively change this conclusion. Finite zigzag and armchair nanoribbons retain optically active channels even when the corresponding two-dimensional bulk response is strongly suppressed. 
In the zigzag ribbon, the perpendicular field reconstructs edge-derived states that are optically dark at weak field into bright photovoltaic channels at strong field by changing their spatial overlap, layer--sublattice composition, and interband
phase structure. 
The inverse-width scaling of the resulting resonance
identifies it as a genuine boundary contribution rather than a remnant of the bulk response. The armchair results further show that magnetic reconstruction of dispersive confined states can also produce large, frequency-selective changes without a pre-existing flat edge band.

These findings suggest practical design principles but also delimit their immediate applicability. The response is not universally enhanced by a magnetic field; instead, its sign and magnitude can be tuned selectively at chosen photon energies through field orientation, edge termination, confinement length, and ribbon width. Structures with a large edge length per illuminated area, such as arrays of narrow ribbons, are natural candidates for amplifying boundary-derived photocurrents. 
The extreme fields used here to expose complete Landau reconstruction are primarily of fundamental interest. Access to analogous regimes at reduced fields may require systems with larger orbital length scales, such as wider confined geometries or graphene superlattices, while the inverse-width scaling imposes a tradeoff between lowering the magnetic crossover scale and retaining a large boundary response. Magnetic proximity in van der Waals heterostructures provides a complementary route to time-reversal-symmetry breaking~\cite{mu2023magnetic}, but it should be distinguished from the
orbital Peierls coupling and Landau quantization considered here.

\section*{Acknowledgments}

\noindent
C.A. and Y.M. acknowledge ANR project COLIBRI No. ANR-22-CE30-0027. 
C.A. acknowledges funding from European Research Council MSCA-ITN TIMES under grant agreement 101118915. 

\bibliography{references.bib}

\end{document}